\documentclass[lettersize,journal]{IEEEtran}
\usepackage{amsmath,amsfonts}
\usepackage{algorithmic}
\usepackage{array}
\usepackage[caption=false,font=small,labelfont=rm,textfont=rm]{subfig}
\usepackage{textcomp}
\usepackage{stfloats}
\usepackage{url}
\usepackage{verbatim}
\usepackage{graphicx}
\def\BibTeX{{\rm B\kern-.05em{\sc i\kern-.025em b}\kern-.08em
		T\kern-.1667em\lower.7ex\hbox{E}\kern-.125emX}}
\usepackage{balance}
\usepackage[utf8]{inputenc}
\usepackage[ruled,vlined]{algorithm2e}
\usepackage{hyperref}

\hypersetup{
	colorlinks = true, 
	linkcolor = blue,
	urlcolor = blue,
	citecolor = blue
}
\usepackage{xcolor}
\begin{document}
	\title{Reconfigurable Intelligent Surface Assisted VEC Based on Multi-Agent Reinforcement Learning}
	\author{Kangwei Qi, Qiong Wu,~\IEEEmembership{Senior Member,~IEEE}, Pingyi Fan,~\IEEEmembership{Senior Member,~IEEE}, \\Nan Cheng,~\IEEEmembership{Senior Member,~IEEE}, Qiang Fan, and Jiangzhou Wang,~\IEEEmembership{Fellow,~IEEE}
	
		\thanks{This work was supported in part by the National Natural Science Foundation of China under Grant No. 61701197, in part by the National Key Research and Development Program of China under Grant No.2021YFA1000500(4), in part by the 111 Project under Grant No. B12018. (Corresponding author: Qiong Wu.)
			
			Kangwei Qi, Qiong Wu are with the School of Internet of Things Engineering, Jiangnan University, Wuxi 214122, China. (e-mail: kangweiqi@stu.jiangnan.edu.cn, qiongwu@jiangnan.edu.cn).
			
			Pingyi Fan is with the Department of Electronic Engineering, Beijing National Research Center for Information Science and Technology, Tsinghua University, Beijing 100084, China (e-mail: fpy@tsinghua.edu.cn).
			
			Nan Cheng is with the State Key Lab. of ISN and School of Telecommunications Engineering, Xidian University, Xi’an 710071, China (e-mail: dr.nan.cheng@ieee.org).
			
			Qiang Fan is with Qualcomm, San Jose, CA 95110, USA (e-mail: qf9898@gmail.com).
			
			Jiangzhou Wang is with the School of Engineering, University of Kent,
			CT2 7NT Canterbury, U.K. (email: j.z.wang@kent.ac.uk).}}
	
	
	\maketitle
	\begin{abstract}
		Vehicular edge computing (VEC) is an emerging technology that enables vehicles to perform high-intensity tasks by executing tasks locally or offloading them to nearby edge devices. However, obstacles may degrade the communications and incur communication interruptions, and thus the vehicle may not meet the requirement for task offloading. Reconfigurable intelligent surfaces (RIS) is introduced to support vehicle communication and provide an alternative communication path. The system performance can be improved by flexibly adjusting the phase-shift of the RIS. For RIS-assisted VEC system where tasks arrive randomly, we design a control scheme that considers offloading power, local power allocation and phase-shift optimization. To solve this non-convex problem, we propose a new deep reinforcement learning (DRL) framework that employs modified multi-agent deep deterministic policy gradient (MADDPG) approach to optimize the power allocation for vehicle users (VUs) and block coordinate descent (BCD) algorithm to optimize the phase-shift of the RIS.	Simulation results show that our proposed scheme outperforms the centralized deep deterministic policy gradient (DDPG) scheme and random scheme.
	\end{abstract}
	

	\section{Introduction}
	\IEEEPARstart{V}EHICULAR edge computing (VEC) is considered to be a promising technology in supporting vehicle real-time computing. It can offload tasks to VEC servers with rich computation resources when the local CPU capacity of the vehicle is limited \cite{r1, r2}. However, in some scenarios, the vehicle user (VU) is impacted by obstacles, making it unable to communicate with base station (BS) over a certain distance \cite{r4}, so that it cannot offload tasks in time.
	
	Recently, reconfigurable intelligent surface (RIS) technology has shown significant advantages in enhancing the communication quality by provisioning an additional communication link to VU by adjusting phase-shifts \cite{r5}, therefore, the VEC system with RIS deserves extensive research. Due to limited hardware, the phase-shift of the RIS can only be selected from a limited number of values \cite{r6}. In addition, the vehicle mobility and uncertain environment pose significant challenges. Therefore, the joint RIS phase-shift and power allocation optimization is a difficult problem to handle. For the optimization problem of RIS phase-shift, in \cite{r7}, He \emph{et al.} employed algorithms such as block coordinate descent (BCD) and alternating optimization (AO) to solve it. In addition, for the VEC power allocation problem, deep reinforcement learning (DRL) algorithms are considered as an effective solution \cite{wu101,wu102,wu103}. In \cite{r8}, Zhu \emph{et al.} used the deep deterministic policy gradient (DDPG) algorithm to allocate the offloading power and local power of a single VU, thereby achieving the minimal total power and buffer length. However, in scenarios with multiple VUs, the centralized algorithms do not have advantages any more \cite{r9}.
	
	In this letter, we investigate the RIS assisted VEC problem and propose a multi-agent DRL scheme with joint BCD\footnote{The source code has been released at: https://github.com/qiongwu86/RIS-VEC-MARL.git}. To simplify the problem, we decompose it into two sub problems: VEC power allocation problem and RIS phase-shift matrix optimization problem. To address it, we first use the BCD algorithm to optimize the RIS phase-shift matrix, and then use the modified multi-agent DDPG (MADDPG) algorithm to solve the power allocation for vehicle offloading and local execution. The extensive simulations demonstrate that the proposed scheme is superior to the centralized DDPG algorithms in terms of rewards, convergence speed, stability, and other aspects.  
	
	\section{System Model and Problem Formulation}\label{System}
	\begin{figure}[t]
		\centering
		\includegraphics[width=2.5in, scale=1.00]{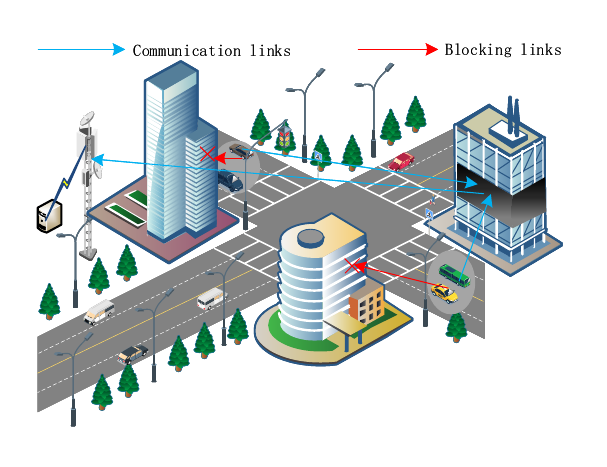}
		\vspace{-0.2cm}
		\caption{RIS aided vehicular edge computing}
		\label{fig1}
		\vspace{-0.5cm}
	\end{figure}
	\subsection{Scenario}
	As shown in Fig. \ref{fig1}, we consider a RIS-assisted VEC network with multiple users, where the BS has $M$ antennas and is associated with a VEC server, and the VU has a single antenna. Since a VU has limited computational resources, it can offload some tasks to the VEC server for processing. When the VU's link is obstructed, it cannot communicate with the BS directly and efficiently, where a RIS with $N$ reflective elements can assist the VU in offloading tasks to the edge devices (we only consider one link, the VU communicating with the BS via the RIS). Here, we consider a urban intersection scenario where VUs are randomly distributed at the intersection. There are paths up and down, left and right, with double lanes of road in each direction and each vehicle randomly chooses a moving direction and at a constant speed. When encountering an intersection, there is a certain probability that the vehicle will choose to turn or go straight ahead. There are $K$ VUs, denoted by the set 
	$\mathcal{K} = \left\{ {1,2, \cdots ,K} \right\}$. The VU has the flexibility to offload the tasks for processing and local execution, and additionally, we divide the task processing into $T$ equally time slots $\Delta t$ of discrete time, denoted by $\mathcal{T} = \left\{ {1,2, \cdots ,T} \right\}$. At each time slot $t$, the vehicle randomly generates a kind of tasks, where task arrivals follow a Poisson distribution with an arrival rate of $\eta $. Note that we consider quasi-static scenarios where the channel conditions keep constant within a time slot but may change between the different time slots.
	
	\vspace{-0.2cm}
	\subsection{Queue Model}	
	For VU $k$, we assume that its task arrival rate at time slot $t$ is 
	${\eta _k}(t)$, which follows a Poisson distribution, so we can calculate the tasks of VU $k$ during time slot t as 
	\begin{equation}\label{eq1}
		{a_k}(t) = {\eta _k}(t) \times \Delta t.
	\end{equation}
	The arriving tasks will be saved in the buffer, and processed at time slot $t+1$, thus we can get the buffer length of VU $k$ at time slot $t+1$ is
	\begin{equation}\label{eq2}
		{q_k}(t + 1) = {\left[ {{q_k}(t) - {q_{k,o}}(t) - {q_{k,l}}(t)} \right]^ + } + {a_k}(t),
	\end{equation}
	where ${q_{k,o}}(t)$ and ${q_{k,l}}(t)$ denote the amount of tasks processed by task offloading and locally, respectively. Note that ${\left[ x \right]^ + } = \max (0,x)$.
	
	\vspace{-0.2cm}
	\subsection{Offloading Model}
	In our proposed model, it is assumed that the location of the $k$th vehicle is $(x_k^t,y_k^t,z_k^t)$ at time slot $t$, and the coordinates of the BS and the RIS are $(B{S_x},B{S_y},B{S_z})$ and $(RI{S_x},RI{S_y},RI{S_z})$, respectively. The link between BS and RIS is line of sight (LoS) and the link between RIS and the vehicles is the same. Therefore, these communication links undergo small-scale fading, which is modeled as the Rician fading with the pure LoS component \cite{r10, r11}. Since both the RIS and the BS are deployed at a fixed location, the RIS-BS link will remain static. Therefore, we can obtain the channel gain ${{\bf{h}}_{r,b}}$ between the RIS and BS as		
	\begin{equation}\label{eq3}
		{{\bf{h}}_{r,b}} = \sqrt {\rho {{\left( {{d_{r,b}}} \right)}^{ - {\alpha _{r,b}}}}} \sqrt {\frac{R}{{1 + R}}} {\bf{h}}_{r,b}^{LoS}.
	\end{equation}
	where $\rho $ is the path loss at reference distance ${d_0} = 1m$, ${d_{r,b}}$ is the geometric distance from the RIS to the BS, ${\alpha_{r,b}}$ is the path loss exponent of the RIS-BS link, and $R$ is the Rician coefficient associated with small-scale fading. LoS components ${\bf{h}}_{r,b}^{LoS}$ is defined as
	\begin{equation}\label{eq4}
		\begin{array}{r}
			{\bf{h}}_{r,b}^{LoS} = [1,{e^{ - j\frac{{2\pi }}{\lambda }{d_{r}}\sin \left( {{\theta _{r,b}}} \right)}}, \cdots ,
			{e^{ - j\frac{{2\pi }}{\lambda }\left( {N - 1} \right){d_{r}}\sin \left( {{\theta _{r,b}}} \right)}}{]^{\rm T}},
		\end{array}
	\end{equation}
	where $\lambda$ is the carrier length, $d_r$ is the interval between RIS elements, and $\theta _{r,b}$ is the departure angle of the signal from the RIS to the BS. Similarly, the channel gain ${\bf{h}}_{k,r}^t$ from the $k$th VU to the RIS at time slot $t$ is defined as
	\begin{equation}\label{eq5}
		{\bf{h}}_{k,r}^t = \sqrt {\rho {{\left( {d_{k,r}^t} \right)}^{ - {\alpha _{k,r}}}}} \sqrt {\frac{R}{{1 + R}}} {\bf{h}}_{k,r}^{t \; LoS},\forall k \in K,\forall t \in T,
	\end{equation}
	where, $d_{k,r}^t$ is the geometric distance between the $k$th vehicle and the RIS at time slot $t$, and ${\alpha _{k,r}}$ is the path loss exponent between the vehicle and the RIS. Note that ${\bf{h}}_{k,r}^{t \; LoS}$ is expressed as
	\begin{equation}\label{eq6}
		\begin{array}{c}
			{\bf{h}}_{k,r}^{t{\rm{ }}LoS} = [1,{e^{ - j\frac{{2\pi }}{\lambda }{d_r}\sin \left( {\theta _{k,r}^t} \right)}}, \cdots ,
			{e^{ - j\frac{{2\pi }}{\lambda }\left( {N - 1} \right){d_r}\sin \left( {\theta _{k,r}^t} \right)}}{]^{\rm T}},
		\end{array}
	\end{equation}
	where ${\theta _{k,r}^t}$ is the arrival angle of the signal from the $k$th vehicle to the RIS at time slot $t$.
	
	In this letter, we consider that the communication link between the vehicle and the BS is completely blocked, and the VU can only communicate through the RIS. Thus, we can obtain the signal-to-noise ratio (SNR) between the $k$th VU and the BS through the RIS at time slot $t$ as
	\begin{equation}\label{eq7}
		{\gamma _k}(t) = \frac{{P_{k,o}^t{{\left| {{{\left( {{{\bf{h}}_{r,b}}} \right)}^H}{\Theta ^t}{\bf{h}}_{k,r}^t} \right|}^2}}}{{{\sigma ^2}}},\forall k \in K,\forall t \in T,
	\end{equation}
	where $P_{k,o}^t\in [0,{P_{\max,o}}]$ is the offloading power of $k$th VU at time slot $t$, and ${\sigma ^2}$ is thermal noise power. The diagonal phase-shift matrix of RIS is ${\Theta ^t} = diag[{\beta _1}{e^{j\theta _1^t}}, \cdots ,{\beta _n}{e^{j\theta _n^t}}, \cdots ,{\beta _N}{e^{j\theta _N^t}}],\forall n \in [1,N]$, ${\beta _n} \in \left[ {0,1} \right]$. Due to hardware constraint, phase-shift can only be selected from a finite discrete value set $\theta _n^t \in \Phi  = \left\{ {0,\frac{{2\pi }}{2^b}, \cdots ,\frac{{2\pi (2^b - 1)}}{2^b}} \right\}$, where $b$ controls the discrete degree of phase-shift.
	
	When VU $k$ chooses to offload tasks to the VEC server associated with the BS, based on the formula \eqref{eq7}, we can obtain the number of offloaded tasks processed by VU $k$ at time slot $t$ as
	\begin{equation}\label{eq8}
		{q_{k,o}}(t) = \Delta t \times W{\log _2}(1 + {\gamma _k}(t)),
	\end{equation}
	where $W$ represents the channel bandwidth.
	
	\vspace{-0.2cm}
	\subsection{Local Execution}
	When VU $k$ selects to process tasks locally, we can obtain the size of tasks that can be processed locally at time slot $t$ as
	\begin{equation}\label{eq9}
		{q_{k,l}}(t) = \Delta t{f_k}(t)/L,
	\end{equation}
	where $L$ is denoted as the CPU frequency required to process one bit of task, 
	${f_k}(t) \in \left[ {0,{F_{\max }}} \right]$ is the CPU frequency scheduled by utilizing DVFS technology to adjust the chip voltage, i.e., 
	\begin{equation}\label{eq10}
		{f_k}(t) = \sqrt[3]{{{p_{k,l}}(t)/c}},
	\end{equation}
	where  ${p_{k,l}}(t) \in [0,{P_{\max ,l}}]$ is the local execution power of VU $k$ at time slot $t$, and $c$ is the effective selection capacitance.
	
	\vspace{-0.2cm}
	\subsection{Problem Formulation}
	The target is  to minimize the power consumption of task offloading and local processing as well as buffer length. Therefore, the multi-objective optimization problem for each VU $k$ can be formulated as follows:
	\begin{subequations}\label{P0}
		\begin{equation}\label{eq11a}
			{P1}:\mathop {\min }\limits_{\theta _n^t,{p_{k,o}}(t),{p_{k,l}}(t)} \left\{ {\frac{1}{T}\sum\limits_{t = 1}^T {\left( {{p_{k,o}}(t) + {p_{k,l}}(t) + {q_k}(t)} \right)} } \right\},
		\end{equation}
		\begin{equation}\label{eq11b}
			{\rm{s}}{\rm{.t}}{\rm{.}}\quad0 < {p_{k,o}}(t) < {P_{\max ,o}},{\rm{ }}\forall k \in \mathcal{K},{\rm{ }}\forall t \in \mathcal{T},
		\end{equation}
		\begin{equation}\label{eq11c}
			0 < {p_{k,l}}(t) < {P_{\max ,l}},{\rm{ }}\forall k \in \mathcal{K},{\rm{ }}\forall t \in \mathcal{T},
		\end{equation}
		\begin{equation}\label{eq11d}
			{\rm{      }}\theta _n^t \in \Phi ,{\rm{            }}\forall n \in \mathcal{N},{\rm{ }}\forall t \in \mathcal{T},
		\end{equation}
	\end{subequations}
	where (\ref{eq11b}) and (\ref{eq11c}) represent the power constraints for VU $k$ when offloading tasks and processing tasks locally, respectively. Due to the limitation of RIS hardware, the size of RIS phase-shift can only be selected within a limited range constrained by (\ref{eq11d}). Furthermore, the objective function is non-convex, so this optimization problem is difficult to be solved. To better address this issue, we propose a multi-agents deep reinforcement learning (MARL) scheme for joint BCD. Since the RIS phase-shift controller and VUs are not the same type of agents, their actions and reward are also different. Therefore, we will first use the BCD algorithm to obtain the optimal size of phase-shift, and then obtain the optimal power allocation scheme through the modified MADDPG algorithm.
	
	\section{Solution Approach}
	We will describe the proposed scheme in detail. Firstly, we calculate the channel information of VU based on its relevant positions to BS and RIS, etc., where the RIS phase-shift is a variable, and we use the BCD algorithm to obtain the optimal phase-shift with the objective of maximizing
	\begin{equation}\label{eq12}
		{\sum\limits_{k = 1}^K {\left| {{{\left( {{h_{r,b}}} \right)}^H}{\Theta ^t}h_{k,r}^t} \right|} ^2},
	\end{equation}
	the details related to the BCD can be referred to \cite{r7}. 
	
	In addition, the power allocation problem of VU is solved according to the modified MADDPG algorithm. Here, each VU interacts with the environment as an agent and makes power allocation decisions through the corresponding strategies. At time slot $t$, the agent obtains the current state $s_t$, the corresponding action $a_t$ and the corresponding reward $r_t$, and transitions to the next state $s_{t+1}$. This process can be formulated as 
	${e_t} = ({s_t},{a_t},{r_t},{s_{t + 1}})$, and the relevant state space, action space, and reward function in this model are represented as follows:
	
	\textbf{State space:} At time slot $t$, the state of each VU $k$ (agent $k$) consists of the following components: buffer length $q_k(t)$, the size of offloaded executed tasks $q_{k,o}(t)$, the size of locally executed tasks $q_{k,l}(t)$, and the offloaded and locally processed task overflows 
	${{q_{k,o}}(t) + {q_{k,l}}(t) - {q_k}(t)}$. In addition, through equation (\ref{eq7}), the SNR of VU $k$ at time slot $t$, i.e. ${\gamma _k}(t - 1)$ , depends on 
	${h_{r,b}}$ and $h_{k,r}^t$, which reflects the channel uncertainty of the VU at time slot $t$. Therefore, the state space of the VU $k$ at slot $t$ can be formulated as ${\bf{s}} = \left[ {{q_k}(t),{q_{k,o}}(t),{q_{k,l}}(t),{q_{k,o}}(t) + {q_{k,l}}(t) - {q_k}(t),{\gamma _k}(t - 1)} \right].	$
	
	\textbf{Action space:} as described above, agent $k$ allocates the offloading power and the local power according to the corresponding policy, so that the action space of agent $k$ at time slot $t$ is defined as 
	${a_k}(t) = \left[ {{p_{k,o}}(t),{p_{k,l}}(t)} \right]$.
	
	\textbf{Reward function:} in our proposed scheme, there are two aspects of rewards for each agent, one is the global reward reflecting the cooperation among agents, and the other is the local reward that is to help each agent to explore to the optimal power allocation scheme. Here, our goal is to optimize the agent $k$'s offloading and the local power level, as well as the corresponding buffer length. To address it, we add two penalties $Pen1$ and $Pen2$, one for the buffer length of agent $k$ being greater than a threshold at time slot $t$, and the other for the penalty incurred when agent $k$ allocates a certain amount of power to offloading execution and local execution at time slot $t$, resulting in the total number of tasks processed being a certain value more than the buffer length. Thus the local and global rewards of VU $k$ at time slot $t$ are	
	\begin{equation}\label{eq13}
		{r_{k,l}} =  - \left[ {{w_1}({p_{k,o}}(t) + {p_{k,l}}(t)) + {w_2}{q_k}(t)} \right] - Pen1 - Pen2,
	\end{equation}
	and \begin{equation}\label{eq14}
		{r_{g}} = \frac{1}{K}\sum\limits_{k \in K} {{r_{k,l}}}.
	\end{equation}
	
	\begin{algorithm}
		\caption{Modified MADDPG Scheme Combined With the BCD Algorithm}\label{al2}
		Start environment simulator, generate vehicles\\
		Initialize global critic networks $Q_{{\psi _1}}^{{g_1}}$ and $Q_{{\psi _2}}^{{g_2}}$\\
		Initialize target global critic networks $Q_{{\psi _1'}}^{{g_1}}$ and $Q_{{\psi _2'}}^{{g_2}}$\\
		Initialize each agent's policy and critic networks\\
		\For{each episode}
		{
			Reset simulation paramaters\\
			\For{each timestep $t$}
			{
				\For{$n = 1,...,N$}
				{
					Fix $n', \forall n' \ne n, n' \in N$ \
					
					Set $\theta _n^t = \mathop {\arg \max }\limits_\Phi  (\ref{eq12})$ \
				}
				
				\For{each agent $k$}
				{
					Observe $s_k^t$ and select action $a_k^t=\pi_\theta(s_k^t)$\\
					Receive local reward $r_{k,l}^t$
				}
				${\bf{s}} = ({s_1},{s_2}, \cdots {s_K})$ and ${\bf{a}} = ({a_1},{a_2}, \cdots {a_K})$ \\
				Receive global reward $r_g^t$\\
				Store $({{\bf{s}}^t},{{\bf{s}}^t},{\bf{r}}_l^t,r_g^t,{{\bf{s}}^{t + 1}})$ in replay buffer $\mathcal{D}$\\
			
			\If{the replay buffer size is larger than $I$}
			{
				Randomly sample mini-batch of $I$ transitions tuples from $\mathcal{D}$\\
				Update global critics by minimizing the loss according to Eq.(\ref{eq18})\\
				Update global target networks parameters: ${\psi _j'} \leftarrow \tau {\psi _j} + (1 - \tau ){\psi _j'}, j=1,2$\\
				
				\If{episode mod $d$}
				{
					\For{each agent $k$}
					{
						Update local critics by minimizing the loss according to Eq.($\ref{eq20}$)\\
						Update local actors according to Eq.($\ref{eq17}$)\\
						Update local target networks parameters: ${\theta _k'} \leftarrow \tau {\theta _k} + (1 - \tau ){\theta _k'}$, ${\phi _k'} \leftarrow \tau {\phi _k} + (1 - \tau ){\phi _k'}$
					}
				}
			}
		}
		}	
	\end{algorithm}
	We propose a MARL algorithm aimed at maximizing both global and local rewards. It utilizes two main critics: a shared global critic evaluates states and actions from all agents to optimize global rewards, while each agent also has a local critic for assessing its own rewards based on local states and actions.
	Each agent has its own actor and critic networks for decision-making and action evaluation. Although each agent can only observe information about the local environment, they can share each other's experience through experience replay. During experience replay, each agent stores and randomly samples from a shared pool of experience, which increases training efficiency and stability while enabling agents to learn how other agents respond to the environment. The global reward encourages collaboration among agents to optimize the system by providing a shared reward signal at each step.
	To better improve the performance of the algorithm, we consider the impact of approximation errors in strategy and value updates on the global critic in MADDPG algorithm, according to \cite{r12}, we employ twin-delay deterministic strategy gradient to replace the global critic.
	
	Specifically, we consider a vehicular environment with $K$ vehicles (agents) and the policies for all agents are $\mathbf{\pi} = \{ {\pi _1},{\pi _2}, \cdots ,{\pi _K}\}$. The agent $k$'s strategy $\pi_k$, Q-functions 
	${\mathop{\rm Q}\nolimits} _{{\phi _k}}^k$ and twin global critic Q-functions 
	(${\mathop{\rm Q}\nolimits} _{\psi 1}^{{g_1}}$, ${\mathop{\rm Q}\nolimits} _{\psi 2}^{{g_2}}$) are parameterized by ${\theta _k}$, $\phi_k$, ${\psi _1}$ and $\psi_2$, respectively. For each agent, the modified policy gradient can be written as
	\begin{equation}\label{eq17}
		\begin{aligned}
			\nabla J({\theta _k}) =& 
			\overbrace {{{\mathbb{E}}_{{\bf{s,a}}\sim \mathcal{D}}}\left[ {{\nabla _{{\theta _k}}}{\pi _k}({a_k}|{s_k}){\nabla _{{a_k}}}Q_{{\psi _j}}^{{g_j}}({\bf{s,a}})} \right]}^{Global\quad Critic} + \\& \underbrace {{{\mathbb{E}}_{{s_k},{a_k}\sim \mathcal{D}}}\left[ {{\nabla _{{\theta _k}}}{\pi _k}({a_k}|{s_k}){\nabla _{{a_k}}}Q_{{\phi _k}}^k({s_k},{a_k})} \right]}_{Local\quad Critic},
		\end{aligned}
	\end{equation}
	where ${\bf{s}} = ({s_1},{s_2}, \cdots {s_K})$ and 
	${\bf{a}} = ({a_1},{a_2}, \cdots {a_K})$ are the total state and action vectors, $\mathcal{D}$ is the replay buffer ,and ${a_k} = {\pi _k}({s_k})$ is the action which is chosen for agent $k$ according to its own policy $\pi_k$. The twin global critic $Q_{{\psi _j}}^{{g_j}}$ is updated to
	\begin{equation}\label{eq18}
		L({\psi _j}) = {{\rm E}_{{\bf{s,a,r,s'}}}}\left[ {{{(Q_{{\psi _j}}^{{g_j}}({\bf{s,a}}) - {y_g})}^2}} \right],{\rm{     }}j = 1,2,
	\end{equation}
	where 
	\begin{equation}\label{eq19}
		{y_g} = {r_g} + \gamma {\mathop {\min }\limits_{j = 1,2}}{\left. {Q_{\psi _j'}^{{g_j}}({\bf{s',a'}})} \right|_{a_k' = \pi _k'(s_k')}}.
	\end{equation}
	Here, ${\bf{\pi}}' = \{ \pi _1',\pi _2', \cdots ,\pi _K'\}$ is the target policy with parameter $\theta ' = \{ \theta _1',\theta _2', \cdots ,\theta _K'\}$. Similarly, the local critic $Q_{{\phi _k}}^k$ of agent $k$ is updated to 
	\begin{equation}\label{eq20}
		{L^k}({\psi _k}) = {{\rm E}_{{s_k},{a_k},{r_k},s_k'}}\left[ {{{(Q_{{\phi _k}}^k({s_k},{a_k}) - y_l^k)}^2}} \right],
	\end{equation}
	where\begin{equation}\label{eq1}
		y_l^k = r_l^k + \gamma {\left. {Q_{\phi _k'}^k(s_k',a_k')} \right|_{a_k' = \pi _k'(s_k')}}.
	\end{equation}
	
	The detailed MARL algorithm is described in the Algorithm \ref{al2}.	
	In addition, Algorithm \ref{al2} contains BCD method and DRL training process, so we separate it into two parts to analyze the computational complexity. The computational complexity of the BCD method is $\mathcal{O}(N2^b)$, its implementation consists of two loops, one is for $N$ RIS elements and one is for $2^b$ optional phase-shift values. Since the modified MADDPG method is implemented in the actor-critic framework, let the computational complexities for calculating gradients and updating parameters in the actor and critic networks being denoted by $G_A$, $G_C$, $U_A$, and $U_C$ respectively. Since the architecture of the target actor network and the target critic network is the same as the network structure of the two mentioned above, their complexities remain the same. Thus, the computational complexity of the modified MADDPG is calculated as
	$\mathcal{O}\left( {{G_A} + {G_C} + 2{U_A} + 2{U_C}} \right) + {\text{ }}\mathcal{O}\left( {2{G_C} + 4{U_C}} \right)$.
	
	\section{Simulation Results}
	\begin{table}[t]
		\caption{Values of the Parameters in the Experiments}
		\vspace{-0.2cm}
		\label{table1}
		\begin{center}
			\begin{tabular}{!{\vrule width1pt}c|c|c|c!{\vrule width1pt}}
				\hline
				
				\hline
				\multicolumn{4}{!{\vrule width1pt}c!{\vrule width1pt}}{Parameters of System Model}\\
				\hline
				
				\hline
				\textbf{Parameter}&{\textbf{Value}}&{\textbf{Parameter}}&{\textbf{Value}} \\
				\hline
				
				\hline
				$K$ & 8 & $\eta$ & 3 Mbps \\
				\hline
				$N$ & 40 & $\sigma^2$ & -110 dBm \\
				\hline
				$b$ & 3 & $\alpha_{r,b}$ & 2.5 \\
				\hline
				$\alpha_{k,r}$ & 2.2 & $W$ & 1 MHz  \\
				\hline
				$L$ & 500 cycles/bit & $c$ & $10^{-28}$ \\
				\hline
				$F_{max}$ & 2.15 GHz & $P_{max,o}, P_{max,l}$ & 1 W\\
				\hline
				
				\hline
				\multicolumn{4}{!{\vrule width1pt}c!{\vrule width1pt}}{Parameters of Modified MADDPG}\\
				\hline
				
				\hline
				\textbf{Parameter} &\textbf{Value} &\textbf{Parameter} &\textbf{Value}\\
				\hline
				
				\hline
				$\alpha_C$ &$0.001$ &$\alpha_A$ &$0.0001$\\
				\hline
				$\omega_{1}$ &$1$ &$\omega_{2}$ &$0.6$\\
				\hline
				$pen1, pen2$ &$2$ &$d$ &$2$\\
				\hline
				$\gamma$ &$0.99$ &$\tau$ &$0.005$\\
				\hline
				$I$ &$64$ &$\mathcal{D}$ &$10^6$\\
				\hline
				
				\hline
			\end{tabular}
		\end{center}
		\vspace{-0.5cm}
	\end{table}
	\begin{figure}[t]
		\centering
		\includegraphics[width=2.5in, scale=1.00]{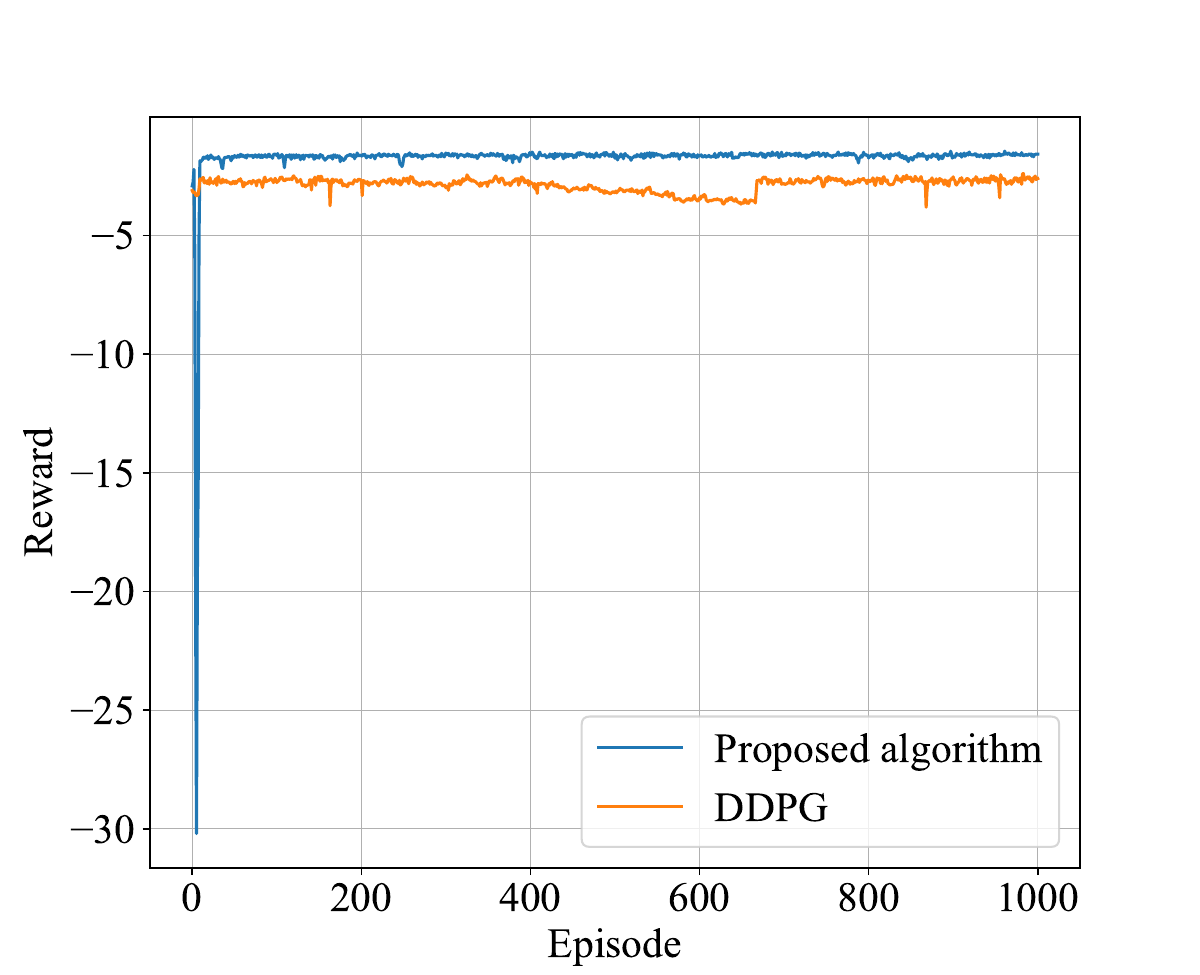}
		\vspace{-0.2cm}
		\caption{Reward function convergence}
		\label{fig2}
		\vspace{-0.5cm}
	\end{figure}
	In this section, we perform simulation experiments to validate our proposed MARL scheme with joint BCD. The simulation tool is Python 3.9. 
	It is assumed that within some road segments, the vehicles cannot communicate directly with the BS $(0,0,25)$ and thus need to rely on the RIS $(220,220,25)$ to communicate indirectly with the BS. $8$ vehicles are involved in the experiment, each with a randomly chosen speed within $10$ to $15$ m/s. The numbers of hidden layers for actor and critic networks are 2 and 3, respectively. 
	The learning rates of critic/actor networks learning rate are 0.001 and 0.0001, respectively. 
	The key parameters are shown in Table \ref{table1}.
	
	
	\begin{figure*}[htbp]
		\centering
		\subfloat[]
		{\includegraphics[width=0.32\textwidth]{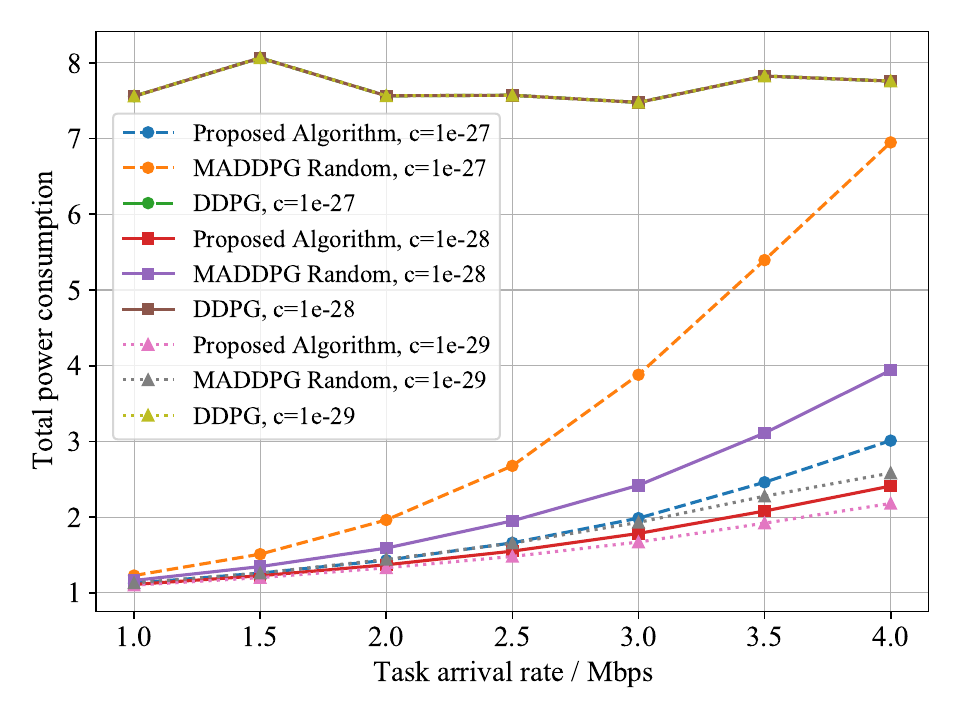}}
		\subfloat[]
		{\includegraphics[width=0.32\textwidth]{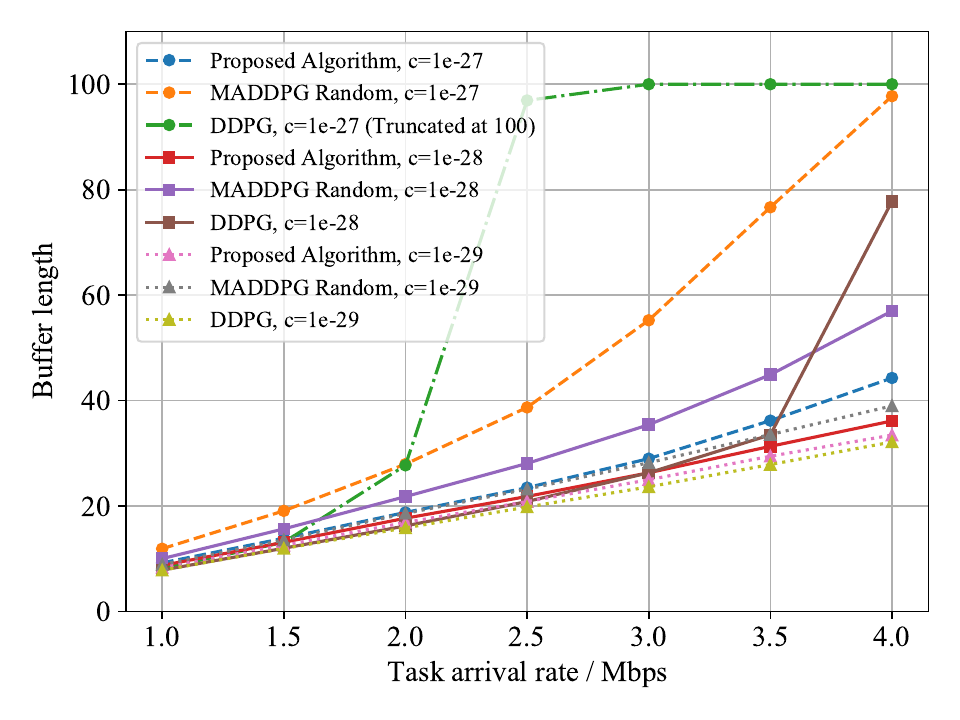}}
		\subfloat[]
		{\includegraphics[width=0.32\textwidth]{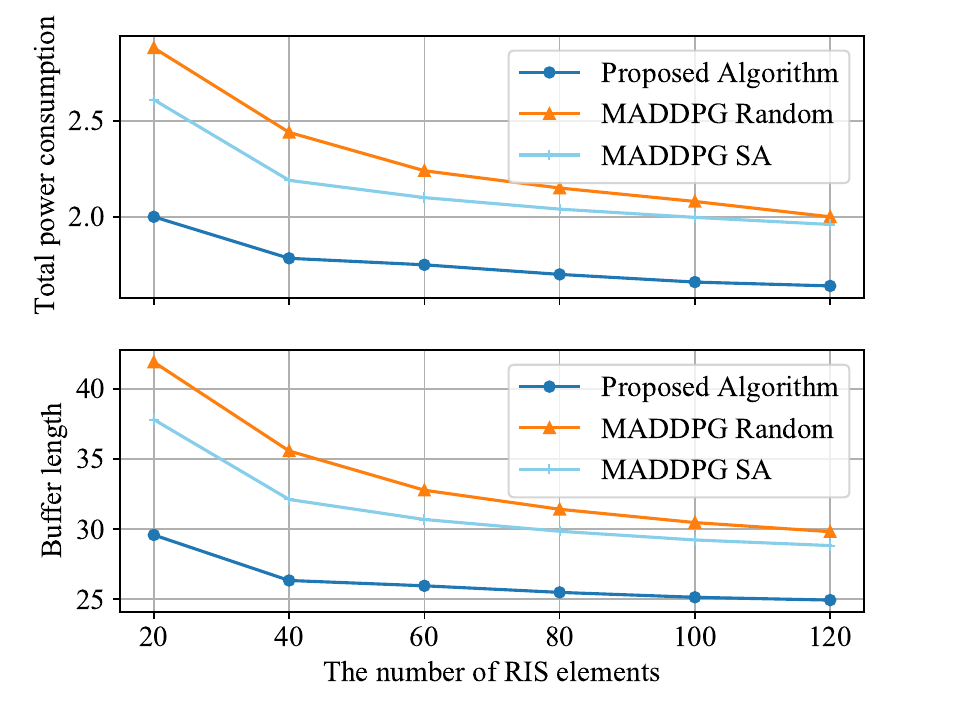}}
		\caption{Performance: (a) Total power consumption; (b) Buffer length; (c) Total power consumption and buffer length}
		\label{fig3}
	\end{figure*}
	
	Fig. \ref{fig2} presents the convergence performance of the proposed scheme and the DDPG scheme in terms of rewards. The proposed framework is very stable with less oscillations compared to DDPG. Since the DDPG is fully centralized, it must take the observations and actions of all agents as input, which cannot effectively address individual performance, leading to suboptimal reward situations. The proposed framework combines MARL with the BCD algorithm and shows excellent convergence performance. It can also simultaneously select the better result on maximizing the local and global rewards of all agents, and thus facilitate the cooperation among agents.
	
	Figs. \ref{fig3}(a) and \ref{fig3}(b) show the changes in power consumption and buffer length for all VUs at different task arrival rates. We also consider different effective selection capacity $c$. It can be seen that the power consumption and buffer length increase with the increment of task arrival rate. The proposed algorithm has the lowest power consumption and relatively small buffer length at different capacitances. As the capacity increases, the power consumption and buffer length also increase. This is because, according to the Eq. (\ref{eq9}) and Eq. (\ref{eq10}), $q_{k,l}(t)$ is related to the effective selection capacity $c$, the larger the capacity, the fewer local processing tasks, which increases the power consumption and buffer length. Due to the DDPG algorithm's poor performance, it uses excess power to manage buffer length. With large capacity and high task arrival rates, it struggles to allocate power effectively, leading to significant task accumulation.
	
	Fig. \ref{fig3}(c) reflects the impact of the number of RIS elements on the network performance, where the total power consumption and buffer length decrease when the number of RIS elements increases. Our proposed scheme outperforms random RIS phase-shift and simulated annealing (SA) phase-shift scheme, because neither of them can be adjusted to the optimal phase-shift, resulting in more power consumption to ensure data transmission and also affecting the buffer length.
	
	\section{Conclusion}
	In this letter, in order to optimize the power consumption, buffer length, and RIS phase-shift matrix in RIS assisted VEC network, we developed a new framework, i.e., the MARL scheme with joint BCD algorithm. Simulation results demonstrated the superiority of our scheme to the random RIS phase-shift and DDPG algorithms in terms of power consumption and buffer length.

\end{document}